\documentclass[12pt,showpacs,aps]{revtex4}

\usepackage{epsfig}

\def\cal#1{{\cal #1}}

\def\m@th{\mathsurround=0pt}
\def\n@space{\nulldelimiterspace=0pt \m@th}
\def\biggg#1{{\mbox{$\left#1\vbox to 20.5pt{}\right.\n@space$}}}

\def\beginenum{\begin{enumerate}}
\def\endenum{\end{enumerate}}
\def\bitem{\begin{itemize}}
\def\eitem{\end{itemize}}
\def\bray{\begin{array}}
\def\eray{\end{array}}
\def\begindoc{\begin{document}}
\def\enddoc{\end{document}}
\def\bq{\begin{equation}}
\def\eq{\end{equation}}
\def\bqy{\begin{eqnarray}}
\def\eqy{\end{eqnarray}}
\def\bqyn{\begin{eqnarray*}}
\def\eqyn{\end{eqnarray*}}
\def\bc{\begin{center}}
\def\ec{\end{center}}
\def\bfll{\begin{flushleft}}
\def\efll{\end{flushleft}}
\def\bflr{\begin{flushright}}
\def\eflr{\end{flushright}}
\newcommand{\Avec}{\mbox{\boldmath $A$}}
\newcommand{\Bvec}{\mbox{\boldmath $B$}}
\newcommand{\Evec}{\mbox{\boldmath $E$}}
\newcommand{\Fvec}{\mbox{\boldmath $F$}}
\newcommand{\Gvec}{\mbox{\boldmath $G$}}
\newcommand{\Rvec}{\mbox{\boldmath $R$}}
\newcommand{\Uvec}{\mbox{\boldmath $U$}}
\newcommand{\Vvec}{\mbox{\boldmath $V$}}
\newcommand{\evec}{\mbox{\boldmath $e$}}
\newcommand{\jvec}{\mbox{\boldmath $j$}}
\newcommand{\kvec}{\mbox{\boldmath $k$}}
\newcommand{\nvec}{\mbox{\boldmath $n$}}
\newcommand{\uvec}{\mbox{\boldmath $u$}}
\newcommand{\vvec}{\mbox{\boldmath $v$}}
\newcommand{\wvec}{\mbox{\boldmath $w$}}
\newcommand{\xvec}{\mbox{\boldmath $x$}}
\newcommand{\omegavec}{\mbox{\boldmath $\omega$}}
\newcommand{\Omegavec}{\mbox{\boldmath $\Omega$}}

\begin{document}

\title{Asymmetry-Driven Structure Formation in Pair Plasmas}
\author{S.M. Mahajan}
\email{mahajan@mail.utexas.edu}
\affiliation{Institute for Fusion Studies, The University of Texas at Austin, Austin,Tx
78712}
\author{N.L. Shatashvili}
\email{shatash@ictp.it}
\author{V.I. Berezhiani}
\affiliation{Andronikashvili Institute of Physics, Tbilisi 0177,
Georgia.\\
Faculty of Exact and Natural Sciences, Department of Physics,
Javakhishvili Tbilisi State University, Tbilisi 0128, Georgia}

\begin{abstract}
{The nonlinear propagation of electromagnetic waves in pair
plasmas, in which the electrostatic potential plays a very
important but subdominant role of a "binding glue" is
investigated. Several mechanisms for structure formation are
investigated, in particular, the "asymmetry" in the initial
temperatures of the constituent species. It is shown that the
temperature asymmetry  leads to a (localizing) nonlinearity  that
is new and qualitatively different from the ones originating in
ambient mass or density difference. The temperature asymmetry
driven  focusing-defocusing nonlinearity supports  stable
localized wave structures in 1-3 dimensions, which, for certain
parameters, may have flat-top shapes.}
\end{abstract}

\maketitle

\clearpage


\section{Introduction}

\label{sec:intro}

The pair plasmas consisting of only positive- and negative-charged
particles of equal mass have attracted special attention mainly
because of the astrophysical applications. In the early universe
during the lepton era, ultra-relativistic electron-positron (e--p)
pairs contribute largely to the matter contents of the Universe
\cite{bib:Wineberg}. The gamma-ray bursts -- the most concentrated
electromagnetic explosions in the Universe -- are believed to be
related with the enormous energy release in compact regions on
short time-scales. This energy release leads to the formation of a
highly dense optically thick e--p plasma that expands and cools
down remaining relativistic \cite{bib:Rees}. Such pair plasmas
exist also in active galactic nuclei, in the relativistic jets,
\cite{bib:blandford}, and  in the pulsar magnetospheres
\cite{bib:sturrock}.

Although there are many  laboratory setups in which e-p pair
plasmas are produced \cite{bib:surko}- \cite{bib:exp-ep}, one of
the most interesting recent laboratory accomplishments is the
successful creation of "sufficiently" dense pair-ion (pi) plasmas
-- the first such plasma consisted of equal-mass, positive and
negative fullerene ions ($C_{60}^+$ and $C_{60}^-$)
\cite{bib:OH1}.

Unlike the e--p plasma systems (both of the astrophysical and
laboratory variety), the fullerene plasma has a long enough life
time that the collective behavior peculiar to the plasma state can
be experimentally investigated under controlled conditions. The
frequencies associated with the collective modes (plasma
frequency, acoustic and Alfv\'en frequencies) in such plasmas tend
to be rather low. Fortunately, the group of Hatekayama and Oohara
have already made considerable progress in the production of the
hydrogen, $H^+$-- $H^-$ plasmas \cite{bib:OH3,bib:OH4}. Since the
initial report, both the quality and quantity of this light pair
ion plasma has been steadily improving \cite{bib:Ha1}.

Since many properties of pair plasmas (a symmetric pair plasma,
for instance, cannot sustain charge separation) are different from
the ordinary electron-ion (e--i) plasma, controlled experiments
would not only advance fundamental physics,  but also create a
laboratory to simulate and understand a variety of phenomena
taking place in astrophysical environments. A basic requirement
for long-time-scale experiments, will be that the pair
annihilation time scale is many orders of magnitude larger than
the plasma period.

Theoretical investigation of pair plasmas have followed two
distinct tracks:

1) the first track emphasizes the special properties that stem
from the symmetric (pair particles have the same inertia,
temperature etc.) nature of the pair plasma. This is a highly
studied field
\cite{bib:shukla-ep,bib:theory-ep,bib:shukla1,bib:shukla2} both in
the  astrophysical as well as the laboratory contexts. In the e-p
plasmas, problems like solitary structure formation are studied
for instance in \cite{bib:TT} while for the ion pair plasmas much
of the linear as well as nonlinear work has been devoted to
understanding and interpreting experimental results (see e.g.
\cite{bib:hans1,bib:frank}), in particular the dispersion curves,
described in \cite{bib:OH1}

2) The second group of papers deals with a whole lot of
interesting phenomena that arise because the symmetry of the pair
plasma is mildly broken through some mechanism which creates some
disparity between the constituent fluids
\cite{bib:H1,bib:H3,bib:shukla4,bib:H5,bib:BM1,bib:BM2,bib:MS,bib:SJK}.
One of the more interesting results  of symmetry breaking is the
creation of localized nonlinear structures; the particular
properties of the structure will, naturally, depend upon the mode
of symmetry breaking.

Although it is natural to imagine that pair plasmas have to be
symmetric: that the charges, densities, temperatures and masses of
positively and negatively charged particles are equal. But both
observations and experiments indicate that asymmetry may appear,
quite naturally, at some stage of their evolution. In fact
asymmetry could also be engineered in experiments if such a state
could show interesting properties. Different species, not produced
in identical conditions \cite{bib:Ha1}, for instance, could have
different thermal speeds (temperatures). One could also arrange
experiments with different setups for different species when, for
instance, there are fractions of heavier/lighter ions or there is
a mixture of different mass/temperature species with opposite
charges. This way one could mimic the conditions pertinent to
astrophysical pair plasmas.

A much investigated  example of broken  symmetry is a pair plasma
contaminated by a small fraction of charged particles with
different mass (lighter/heavier than main species). Symmetry
breaking could also occur when the constituent elements  of the
two fluids have slightly different masses, or the  fluids have
slightly different temperatures. As expected, symmetry breaking
induces new properties  that are different from what pertains for
pure pair plasmas. It was shown in \cite{bib:BM1,bib:BM2,bib:MS}
that pair plasma, contaminated by the heavier immobile ions, can
support 3D stable completely localized structures of EM radiation
-- "light- bullets", non-diffracting and nondisspersive EM pulses
of pancake shape with large density bunching. Existence of such
localized structures is not possible in  pure pair plasmas.
Localized nonlinear structures of EM radiation were also found in
a hot e-p relativistic plasma containing a small fraction of cold
electron---ion component \cite{bib:SJK}. Similar behaviour could
be expected in doped (or dust-contaminated) fullerene plasmas in
laboratory \cite{bib:shukla3}.

The present study concentrates on establishing the existence of
electromagnetic (EM) solitonic structures in pair plasmas that
could have been created due to asymmetries of different origin. We
will work out the consequences of two classes of symmetry
breaking; asymmetry arising from a small temperature difference of
the constituent species and the asymmetry arising from a small
difference in the species masses. The plasma is assumed to be
underdense while the EM pulse is longer then characteristic skin
length of the plasma. In electron-ion underdense plasma the
formation of solitonic structures takes place only at the
frequency close to the plasma frequency since in such plasma Raman
instabilities dominate the process of soliton formation. In
contrast, in pure symmetric pair plasma, ponderomotive forces are
same for different species and the excitation of longitudinal
waves by the EM pulse and Raman instabilities can not develop. In
slightly asymmetric pair plasmas the ponderomotive forces acting
on positively and negatively charged species are slightly
different; the generation of weak ambipolar electrostatic
potential takes place and the effects related to Raman
instabilities can be ignored to leading order. This potential
plays a fundamental role in structure formation; it acts as the
binding "glue" that concentrates matter and radiation in a small
region. The nonlinearity due to temperature asymmetry is found to
have a focussing-defocussing form (different from the
nonlinearities originating in other modes of symmetry breaking)
imparting a rich structure to the corresponding solitons.

\section{Model}

\label{sec:model}

\bigskip

Let us assume that the velocity distribution of particles is locally a
relativistic Maxwellian. Then the dynamics of the fluid of species $\alpha$ (%
$\alpha$ represents negative and positive particles of any origin)
is contained in the equations (see for details \cite{bib:BMYO}):
\begin{equation}
{\frac{\partial}{\partial
t}}(G_{\alpha}\mathbf{p}_{\alpha})+m_{0\alpha}c^2 \nabla
(G_{\alpha}\gamma_{\alpha})=e_{\alpha}\mathbf{E}+(\mathbf{u}_{\alpha}\times
\mathbf{\Omega_{\alpha}}) , \label{eq:H-00}
\end{equation}
whose \ {\it curl} \ converts the equation of motion to the
vortex-dynamical form:
\begin{equation}
{\frac{\partial \mathbf{\Omega_{\alpha}}}{\partial t}}=\nabla \times (%
\mathbf{u}_{\alpha}\times \mathbf{\Omega_{\alpha}}),  \label{eq:H-0}
\end{equation}
where \
$\mathbf{p}_{\alpha}=\gamma_{\alpha}m_{\alpha}\mathbf{u}_{\alpha}$
\  is the hydrodynamic momentum, \ $\mathbf{E}$ and $\mathbf{B}$ \
are the electric and magnetic fields and \
$\mathbf{\Omega_{\alpha}}=(e_{\alpha}/c)\mathbf{B} +\nabla \times
G_{\alpha}\mathbf{p}_{\alpha}$ \ is the so called generalized
vorticity. Here \ $\mathbf{u}_{\alpha}$ \ denotes he hydrodynamic
velocity, \
$\gamma_{\alpha}=(1-\mathbf{u}_{\alpha}^2/c^2)^{-1/2}=(1+
\mathbf{p}_{\alpha}^2 /m_{0\alpha}^2c^2)^{1/2}$ \ is the
relativistic factor and \
$m_{\alpha}\,G_{\alpha}(z_{\alpha})=m_{\alpha}\,K_3
(z_{\alpha})/K_2 (z_{\alpha})$ \ is the thermally enhanced
"effective mass", [$z_{\alpha} = m_{\alpha} c^2/T_{\alpha}$],
where \ $K_{\nu}$ \ are the modified Bessel functions and \
$m_{\alpha}$ \ and \ $T_{\alpha}$ \ are the particle rest mass and
temperature of species $\alpha$, respectively. For nonrelativistic
temperatures \ ($T_{\alpha}\ll m_{\alpha}c^2$) \
$G_{\alpha}=1+5T{\alpha}/2m_{\alpha}c^2$ \ and for the
ultra-relativistic temperatures ($T_{\alpha}\gg m_{\alpha}c^2$) \
$G_{\alpha}= 4T_{\alpha}/m_{\alpha}c^2\gg 1$. Note, that the
relativistic thermal pressure $\mathit{P}_{\alpha}$ \
[$=(n_{\alpha}/ \gamma_{\alpha})T_{\alpha}$ in the rest frame with
\ $n_{\alpha}$\ being the density in the laboratory frame of the
pair-fluid] \ appears through the temperature dependent factor \
$G_{\alpha}$ \ defined by \ $\gamma_{\alpha}\nabla
\mathit{P_{\alpha}}= m_{\alpha}c^2n_{\alpha}\nabla G_{\alpha}$.
The system of Eqs. (\ref{eq:H-00})-(\ref{eq:H-0}) is augmented by
the equation of state:
\begin{equation}
\frac{n_{\alpha}z_{\alpha}}{\gamma_{\alpha}K_2(z_{\alpha})}\
\mathrm{{exp}[-z_{\alpha}K_2(z_{\alpha})]=const_{\alpha} ,}
\label{eq:H-01}
\end{equation}
which yields the usual results ($n_{\alpha}^{^{\prime
}}T_{\alpha}^{3/2}=const $ \ in non-relativistic limit for mono-atomic gas
and $n_{\alpha}^{^{\prime }}T_{\alpha}^3=const $ in ultrarelativistic case
for photons).

From (\ref{eq:H-0}) it follows, that if the generalized vorticity
is initially zero ($\mathbf{\Omega_{\alpha}}=0$) everywhere in
space, it will remain zero for all subsequent times. We assume
that before the EM radiation is "switched on" the generalized
vorticity of the system is zero.

And for both species we have the continuity equation:
\begin{equation}
{\frac{\partial n_{\alpha}}{\partial t}}+\nabla \cdot (n_{\alpha}
\mathbf{u}_{\alpha})=0.  \label{eq:H-4}
\end{equation}

To study the nonlinear propagation of intense EM wave in a pair
plasma (could be relativistically hot) consisting of negative and
positive ions we must couple the equations of motion with Maxwell
equations. In terms of the vector \,($\mathbf{A}$)\, and
electrostatic \,($\phi$)\, potentials defined by:
\begin{equation}
\mathbf{E}=-{\frac{1}{c}}{\frac{\partial\mathbf{A}}{\partial t}}- \nabla
\varphi ;\ \ \ \ \ \mathbf{B}= \nabla\times \mathbf{A},  \label{eq:H-5}
\end{equation}
the basic equations take the form (Coulomb gauge $\nabla\cdot \mathbf{A}=0$):
\begin{equation}
{\frac{\partial^2 \mathbf{A}}{\partial t^2}}-c^2 \triangle \mathbf{A}+c{%
\frac{\partial}{\partial t}}(\nabla \varphi)-4\pi c\mathbf{J}=0,
\label{eq:H-6}
\end{equation}
and
\begin{equation}
\triangle \varphi=-4\pi\rho,  \label{eq:H-7}
\end{equation}
where for the charge and current densities we have respectively:
\begin{equation}
\rho=\sum_{\alpha}e_{\alpha}n_{\alpha} ;\ \ \ \ \ \ \mathbf{J}%
=\sum_{\alpha}e_{\alpha}n_{\alpha}\mathbf{u}_{\alpha} .  \label{eq:H-8}
\end{equation}

Our purpose is to explore the possibility of finding localized
structures in such plasmas; the equilibrium state is characterized
by charge neutrality: $n_0^+=n_0^-$, where $n_0^+$ and $n_0^-$ are
the positive and negative charge unperturbed densities. The
subscript $\alpha$ hereafter will indicate the negative
($\alpha=-$) and the positive ($\alpha=+ $) ions. In terms of
dimensionless variables:
\[
\mathbf{p}^{\pm}=\frac{\mathbf{p}^{\pm}}{m^-c}\ , \quad
n^{\pm}=\frac{n^{\pm}}{n_{0}^{\pm}}\ , \quad
T^{\pm}=\frac{T^{\pm}}{m^-c^2} , \quad \mathbf{A}=
\frac{|e|\mathbf{A}}{m^-c^2}\ ,
\]
\begin{equation}
\phi=\frac{|e|\varphi}{m^-c^2}\ , \qquad
\mathbf{r}=\frac{\omega_-}{c}\mathbf{r}\ , \qquad t=\omega_-t \ ,
\label{eq:H-10}
\end{equation}
where $\omega_-=(4\pi e^2n_0^-/m^-)^{1/2}$ is the Langmuir frequency of
negative species, the entire set of defining equations reads:
\begin{equation}
{\frac{\partial^2 \mathbf{A}}{\partial t^2}}-\triangle
\mathbf{A}+{\frac{\partial}{\partial t}}(\nabla
\phi)+\left[\frac{n^-\mathbf{\Pi}^-}{\Gamma^-}
-\frac{n^+\mathbf{\Pi}^+}{\Gamma^+}\right]=0,  \label{eq:H-11}
\end{equation}
\begin{equation}
\triangle \phi= n^--n^+ ,  \label{eq:H-12}
\end{equation}
\begin{equation}
\frac{\partial}{\partial t}\Pi^{\pm}+\nabla \Gamma^{\pm} = \mp
\frac{\partial \mathbf{A}}{\partial t}\mp\nabla\phi ,
\label{eq:H-14}
\end{equation}
\begin{equation}
{\frac{\partial n^{\pm}}{\partial t}}+\nabla \cdot \left(n^{\pm}
\frac{\mathbf{\Pi}^{\pm}}{\Gamma^{\pm}}\right)=0 , \label{eq:H-15}
\end{equation}
where it was convinient to introduce temperature dependent
momentum $\mathbf{\Pi}^{\pm}=G^{\pm}\mathbf{p}^{\pm}$ and
relativistic factor
$\Gamma^{\pm}=G^{\pm}{\gamma}^{\pm}=\sqrt{(G^{\pm})^2+(\mathbf{\Pi}^{\pm})^2}$.
The equations of state for species then read:
\begin{equation}
\frac{n^{\pm}}{\Gamma^{\pm}f(T^{\pm})}=\frac{1}{G^{\pm}_{\infty}f(T^{\pm}_{\infty})}
\label{eq:H-16}
\end{equation}
with $f(T^{\pm})=[T^{\pm}K_2\left(1/T^{\pm}\right)/G^{\pm}]\, \mathrm{{exp}
[ G^{\pm}/T^{\pm}]}$ and $T^{\pm}_{\infty}$ the equilibrium temperature of
species.

\bigskip

We now discuss the propagation (along the $z$ axis) of a
circularly  polarized EM wave with a mean frequency $\omega_o$ and
a mean wave number $k_o$ \begin{equation}
\mathbf{A}_{\perp}={\frac{1}{2}}(\mathbf{x}+i\mathbf{y})A(\mathbf{r}%
_{\perp},z,t) exp(ik_o z-i\omega_o t) +c.c. ,  \label{eq:H-18}
\end{equation}
where $A(z,t)$ is a slowly varying function of $z$ and $t$ and
$\mathbf{x}$ and $\mathbf{y}$ are the standard unit vectors. The
choice of circular polarization is dictated by considerations of
simplicity (prevents harmonic generation). We shall now follow
standard methods to analyze the system. We assume that
the longitudinal extent of the pulse is much shorter than its
transverse dimensions ($\partial A/\partial z \gg \nabla _{\perp
}A$). The gauge condition gives us
$A_z=(i/k_0)(\nabla_{\perp}\cdot \mathbf{A}_{\perp}); \ |A_z|\ll
|A_{\perp}|$. Consequently the effects related to $A_z$ will turn
out to be negligibly small. Then, in the slowly varying amplitude
approximation, the transverse component of (\ref{eq:H-14}) is
integrated yielding:
\begin{equation}
\mathbf{\Pi}_{\perp}^{\pm}=\mp\mathbf{A}_{\perp},  \label{eq:H-19}
\end{equation}
where the constant of integration is set equal to zero since particle
hydrodynamic moments are assumed to be zero at infinity where the field
vanishes. Note that for the longitudinal motion the equations of motion can be
treated one-dimensionally.

Longitudinal dynamics is described by the $z$ component of the
equation of motion (\ref{eq:H-14}) and continuity equation. We
note, that due to the circular polarization of EM wave
$\gamma^{\pm}=[1+|A|^2/(G^{\pm})^2+(\Pi_z^{\pm})^2/(G^{\pm})^2]^{1/2}$
doesn't depend on the fast time $\omega_0^{-1}$ and all the
quantities in (\ref{eq:H-14}) vary on a slow time scale.
Therefore, we can introduce the following variables for
convenience: $\xi =z-v_gt \ , \tau =t$,\ where $v_g=k_0/\omega_0$
\ is the group velocity of the EM wave packet. Assuming
$v_g\partial/\partial \xi\gg
\partial/\partial \tau$, straightforward algebra gives the following
integral of motion:
\begin{equation}
G^{\pm}\left[1+\frac{|A|^2}{(G^{\pm})^2}+\frac{(\Pi_z^{\pm})^2}{(G^{\pm})^2} %
\right]^{1/2}-v_g\Pi_z\pm \phi =const^{\pm} .  \label{eq:H-23}
\end{equation}
The constants of integration are determined from the boundary conditions:
the EM fields and plasma momenta vanish at infinity; they are found to
be $G_{\infty}^{\pm}(T_{\infty}^{\pm})$.

In this paper we will deal with transparent plasmas, i.e,
$\omega_0\gg 1$, and $v_g\simeq1$; such plasmas are found both in
astrophysical and laboratory conditions (non-transparent plasma
case will be discussed in a future publicarion). From the
continuity equation we have:
\begin{equation}
n^{\pm}=\frac{\gamma^{\pm}}{\gamma^{\pm}-p_z}  \label{eq:H-24}
\end{equation}
and straightforward algebra leads to:
\begin{equation}
\gamma^{\pm}-p_z=\frac{G_{\infty}^{\pm}}{G^{\pm}}\left[1\mp\frac{\phi} {%
G_{\infty}^{\pm}}\right]^{-1} ,  \label{eq:H-25}
\end{equation}
\begin{equation}
\frac{n^{\pm}}{\gamma^{\pm}}=\frac{G^{\pm}}{G_{\infty}^{\pm}}\left[1\mp\frac{%
\phi} {G_{\infty}^{\pm}}\right]^{-1} ,  \label{eq:H-26}
\end{equation}
where $G_{\infty}^{\pm}\equiv G_{\infty}^{\pm}(T_{\infty}^{\pm})$. Equation (\ref%
{eq:H-23}) yields:
\begin{equation}
p_z^{\pm}=\frac{G^{\pm}}{2G_{\infty}^{\pm}}\left[1+\frac{|\mathbf{A}|^2 }{%
(G^{\pm})^2}-\frac{G_{\infty}^{\pm}}{(G^{\pm})^2}\left[1\mp\frac{\phi} {%
G_{\infty}^{\pm}}\right]^2\right]\left[1\mp\frac{\phi} {G_{\infty}^{\pm}}%
\right]^{-1}  \label{eq:H-27}
\end{equation}
and
\begin{equation}
\gamma^{\pm}=\frac{G^{\pm}}{2G_{\infty}^{\pm}}\left[1+\frac{|\mathbf{A}|^2 }{%
(G^{\pm})^2}+\frac{G_{\infty}^{\pm}}{(G^{\pm})^2}\left[1\mp\frac{\phi} {%
G_{\infty}^{\pm}}\right]^2\right]\left[1\mp\frac{\phi} {G_{\infty}^{\pm}}%
\right]^{-1}  \label{eq:H-28}
\end{equation}
which allow us to write densities fully in terms of potentials $\phi$ and $%
\mathbf{A}$:
\begin{equation}
n^{\pm}=\frac{G^{\pm}}{2G_{\infty}^{\pm}}\left[\left(1+\frac{|\mathbf{A}|^2
}{(G^{\pm})^2}\right)\left(1\mp\frac{\phi} {G_{\infty}^{\pm}}\right)^{-2}+1%
\right].  \label{eq:H-29}
\end{equation}

In pure pair plasma with equal temperature species, the radiation
pressure gives equal longitudinal momenta to both the negative and
positive ions (since their effective masses are equal
[$G^-=G^+=G$]) and thus does not produce any charge separation
($n_-=n_+$ and $\phi=0$) (Berezhiani and Mahajan, 1994).
Berezhiani and Mahajan (1995) showed that the situation changes by
introducing a small fraction of heavy ions; with the  "symmetry
breaking" between hot electrons and positrons, it becomes possible
to generate a finite $\phi$. Several studies showed the existence
of electrostatic potential due to the small fraction of different
temperature electrons \cite{bib:H1,bib:shukla4}; pair plasmas
respond similarly \cite{bib:SJK}.

In \cite{bib:MS} it was demonstrated that in the  pair ion
plasmas, for which the symmetry is  broken by a slight
contamination (doping) through a heavier immobile ion, the
electrostatic potential is no longer zero and such plasmas can
support stable localized EM wave structures even in the
nonrelativistic limit appropriate to the current and near future
laboratory experiments. As we will see below, the creation of
electrostatic potential is also possible due to the difference in
initial temperatures for different (major constituent) species
(yielding the difference in "effective masses" of species).

\section{Formation of Localized Structures in Pair Plasmas with Temperature
Asymmetry}

We introduce the "asymmetry" through temperature difference
between the two species. The existence of temperature asymmetry
is both experimentally and  observationally justified (see the
references given in the Introduction) and leads to different
"effective masses" even though the real masses are equal
($m^-=m^+$). We would expect that this asymmetry in "effective
masses" would lead to effects similar to the ones studied for
ordinary e-i plasmas. As a first step we assume, that temperatures
are only slightly different
\begin{equation}
\frac{G_{\infty}^-}{G_{\infty}^+}= \beta \neq 1 ,  \label{eq:M-1}
\end{equation}
$\beta $ is close to $1$ so that we can have small but finite $\phi$.

This assumption lets Eqs. (\ref{eq:H-26}) to be written as
\begin{equation}
\frac{n^+}{\gamma^{+}}=\frac{G^+}{G_{\infty}^+}\left[{1-\frac{\phi}
{G_{\infty}^+}}\right]^{-1}=\frac{g^+}{1-\beta\hat{\phi}} \ ,
\label{eq:M-2}
\end{equation}
\begin{equation}
\frac{n^-}{\gamma^{-}}=\frac{G^-}{G_{\infty}^-}\left[{1+\frac{\phi}
{G_{\infty}^-}}\right]^{-1}=\frac{g^-}{1+\hat{\phi}},
\label{eq:M-3}
\end{equation}
where the following normalization
$g^{\pm}=(G^{\pm}/G_{\infty}^{\pm})$ and
$\hat{\phi}=\phi/G_{\infty}^{-}$ were used. Eq. (\ref{eq:H-29}),
written fully  in terms of potentials $\phi , A$, relativistic
factors $G^{\pm}$ and $\beta$, transforms to:
\[
n^{+}=\frac{1}{2}\left[1+{\frac{(g^+)^2+\beta^2|\hat{A}|^2}
{(1-\beta\hat{\phi})^2}}\right] ,
\]
\begin{equation}
\
n^{-}=\frac{1}{2}\left[1+{\frac{(g^-)^2+|\hat{A}|^2}{(1+\hat{\phi})^2}}\right],
\label{eq:M-5}
\end{equation}
where $\hat{A}=A/G_{\infty}^-$. From now on we will omit the \
${\hat{..}}$ \ and it will be assumed that $\phi $ and $A$ are
normalized to $G_{\infty}^- $.

For clarity of exposition we will not attempt to analyze the
general case  confining ourselves to the two extreme limits:
non-relativistic and super-relativistic temperatures.

To close  Maxwell equations, we need to evaluate
$\mathbf{J}^{\pm}=n^{\pm}\mathbf{\Pi}^{\pm}/\Gamma^{\pm} $, the
current that goes into (\ref{eq:H-11}). Using  Eqs. equations
(\ref{eq:M-2}), (\ref{eq:M-3})  and the normalized version of  Eq.
(\ref{eq:H-19})
\begin{equation}
g^{\pm}\mathbf{p}_{\perp}^{\pm}=\mp \mathbf{A} \ .
\label{eq:M-12}
\end{equation}
the expression for the total current is easily derived  to be:
\begin{equation}
\mathbf{J}_{\perp}= -\left[{\frac{1}{1-\beta \,\phi
}}+\frac{1}{1+\phi }\right]\,\mathbf{A} \ .  \label{eq:M-13}
\end{equation}
Using the quasi-neutrality condition $n^+=n^-$ (characteristic
length-scale of wave $L\gg 1$) and plugging it into the equations
(\ref{eq:M-5}), and taking into account that for a transparent
plasma, heating of both fluids is very weak (implying $g^{\pm}\sim
1$) we can derive $\phi \sim (1-\beta) \,\psi (|A|^2)$ (where
$\psi(|A|^2)\leq 1$). Putting all the  pieces together (evaluating
the current, actually the $(\mathbf{J}_{\perp}-2\mathbf{A})$ in
our notation) the nonlinear term in Eq. (\ref{eq:H-11}) is
expressible as:
\begin{equation}
\mathrm{NL}=\frac{1}{1-\beta \,\phi }+\frac{1}{1+\phi }-2
\simeq-\phi\,[(1-\beta)-2\,\beta\,\phi] ,  \label{eq:M-39}
\end{equation}
where we do not neglect $\phi$ \ w.r.t. $(1-\beta)\ll 1$.  To
complete Maxwell's equations, we have to relate the electrostatic
potential $\phi$ ($<1$) with its source, in this case the
temperature difference between the species.

\subsection{Super-relativistic temperature pair plasmas}

For super-relativistic temperatures $T^{\pm}\gg 1$ and $%
G^{\pm}=4T^{\pm}(\gg 1)$, $g^{\pm}=T^{\pm}/T_{\infty}^{\pm}$,
and Eq. (\ref{eq:H-16}) reads:
\begin{equation}
\frac{n^{\pm}}{\gamma^{\pm}}=\left(\frac{T^{\pm}}{T_{\infty}^{\pm}}
\right)^3 ={g^{\pm}}^3.  \label{eq:M-8}
\end{equation}
which, in conjunction with (\ref{eq:M-2}), (\ref{eq:M-3}) and
(\ref{eq:M-8}),  yields :
\begin{equation}
g^+=\frac{1}{(1-\beta \phi )^{1/2}} \ , \qquad \qquad
g^-=\frac{1}{(1+\phi)^{1/2}}  \label{eq:M-10}
\end{equation}
evaluating the "effective mass" \ $G^{\pm}$ self--consistently in
terms of $\phi $. Straightforward algebra and the use of
quasi-neutrality condition yields
\[
\qquad \qquad \phi \simeq \frac{|A|^2}{3}\,[(1-\beta)-2\,\beta\,\phi]
\]
leading, finally, to
\begin{equation}
\phi = \frac{(1-\beta)}{2}\ \frac{\kappa |A|{\,^2}}{\left({1+\beta\,\kappa
|A|^{\,2}}\right)} \quad \mathrm{with} \quad \kappa \equiv \frac{2}{3} \ .
\label{eq:M-38}
\end{equation}
We see that $\phi \sim (1-\beta)$ when $|A|^2\gg 1$ and $\phi \leq(1-\beta)$
when $|A|^2\leq 1$ and our estimation of nonlinear term (\ref{eq:M-39})
remains valid.

Note, that with electrostatic potential defined by (\ref{eq:M-38}) one learns that
that heating/cooling of both fluids is weak ($g^+\sim (1+0.5\beta
\,\phi); \ g^-\sim (1-0.5\,\phi)$). Also from the Eqs.(\ref{eq:M-5}) the
species densities come out to be:
\begin{equation}
n^{\pm}\sim 1+\frac{\beta}{2}|A|^2 .  \label{eq:M-40}
\end{equation}

\subsection{Non-relativistic temperature pair plasmas}

For non-relativistic temperatures $T^{\pm}, T_{\infty}^{\pm}\ll 1$
and $G^{\pm}=1+5T^{\pm}/2$ .  In this limit the relevant relations
are:
\begin{equation}
\frac{n^{\pm}}{\gamma^{\pm}}=\left(\frac{T^{\pm}}{T_{\infty}^{\pm}}\right)^{3/2}
, \quad g^{\pm} \simeq 1+\frac{5}{2}\,(T^{\pm}-T_{\infty}^{\pm}),
\label{eq:M-26}
\end{equation}
and
\begin{equation}
\beta = \frac{G_{\infty}^-}{G_{\infty}^+} \simeq 1+\frac{5}{2}(T_{\infty
}^{-}-T_{\infty}^+).  \label{eq:M-27}
\end{equation}
Straightforward but tedious algebra  leads us to
\begin{equation}
g_{\pm}=1+H_{\pm},  \label{eq:M-28}
\end{equation}
where
\begin{equation}
H_+=\frac{5}{2}\,\beta \,T_{\infty}^+\,\phi \qquad \mathrm{and}
\qquad H_-=-\frac{5}{2}\,T_{\infty}^-\,\phi \ .  \label{eq:M-29}
\end{equation}
From (\ref{eq:M-28})--(\ref{eq:M-29}) we find:
\[
g_++g_- \simeq 2+(\beta-1)\,\phi \ ,
\]
\begin{equation}
g_+-g_- \simeq \frac{5}{2}\,\phi \,T_{\infty}^-(1+\beta ) ,  \label{eq:M-41}
\end{equation}
from which
\begin{equation}
g_{+}^2-g_{-}^2 \simeq \,\phi T_{\infty}^{-}(1+\beta ) ,  \label{eq:M-32}
\end{equation}
readily follows. Repeating the same procedure as we adopted earlier, we may derive
\begin{equation}
\phi = \frac{(1-\beta)}{2}\ \frac{\kappa |A|{\,^2}}{\left({1+\beta\,\kappa
|A|^{\,2}}\right)} , \quad \mathrm{with} \quad \kappa \equiv \frac{1}{2}\
\label{eq:M-43}
\end{equation}
an expression that has the same general form as
Eq.(\ref{eq:M-38}) that pertains to the super-relativistic
temperature regime! In fact,  the two limits are very similar --
for either case the plasma heating/cooling is weak and the density
bunching $n^{\pm}\sim 1+\beta \,|A|^2/2$.

\section{Localized Structures}

Our interest, here, is  to delineate the conditions for the
existence of localized structures in pair plasmas. For the
temperature asymmetry driven system, the perpendicular current is
assembled from Eqs. (\ref{eq:M-38}) and (\ref{eq:M-43}):
\[
J_{\perp }\simeq -(2-\phi \,[\,(1-\beta )-2\beta \,\phi ]\,)\ A=
\]%
\begin{equation}
=-\left[ 2-\frac{(1-\beta )^{2}}{2}\frac{\kappa
\,|A|{\,^{2}}}{\left( {1+\beta \,\kappa \,|A|^{\,2}}\right)
^{\,2}}\right] \,A  \label{eq:S-4}
\end{equation}%
with \ $\kappa =1/2$ \ for non-relativistic temperatures and \
$\kappa =2/3$ \ for relativistic temperatures. The final equation
for the vector potential (equivalent to the final defining
equation (43) of \cite{bib:BM2}), derived from  Eq.
(\ref{eq:H-11}) and the preceding expression of current, reads:
\[
2i\omega _{0}\frac{\partial A}{\partial \tau }+\frac{1+\beta }{\omega
_{0}^{2}}\ \frac{\partial ^{\,2}A}{\partial \xi ^{2}}+\nabla _{\perp
}^{2}\,A+
\]
\begin{equation}
+\frac{(1-\beta )^{2}}{2}\,\frac{\kappa \,|A|{\,^{2}}}{\left( {1+\beta
\,\kappa \,|A|^{\,2}}\right) ^{2}}\,A=0,  \label{eq:S-5}
\end{equation}
\[
\mathrm{with}\qquad \epsilon ^{2}\equiv \frac{1}{4}(1-\beta )^{2}\ll 1,
\]
where we have redefined \ $m_-$ \ as \ $m_-\to m_-G_{\infty}^{-}$
\  that introduces the "effective mass" for negatively charged
ions. The wave frequency \ $\omega_0$ \ satisfies the dispersion
relation: \ $\omega_0^2=k_0^2+(1+\beta)$ \ (in dimensional units
this reads as: \ $\omega_0^2=k_0^2\,c^2+(1+\beta)\ \omega_-^2$ ) .
In Eq. (\ref{eq:S-5}) the weak dependence on the transverse
coordinates has been retained. Note, that in spite of the fact
that $\partial A/\partial \xi \gg \nabla _{\perp }A$ the second
and the third terms can be comparable because of the "transparent
plasma" ($\omega_0^{2}\gg 2$) condition \cite{bib:Chen}.

With self-evident renormalization, Eq.(\ref{eq:S-5}) can
be written as:
\begin{equation}
i\frac{\partial A}{\partial \tau }+\frac{\partial^{\,2}A}{\partial
\xi ^{2}} +\nabla _{\perp }^{2}\,A+F(|A|^{2})\cdot A=0,
\label{eq:S-6}
\end{equation}%
with the nonlinearity function given by:
\begin{equation}
F(|A|^{2})=\frac{|A|^{2}}{(1+|A|^{2})^{2}}\ .  \label{eq:S-7}
\end{equation}%
Equation (\ref{eq:S-6}) is nothing but the nonlinear
Schr\"{o}dinger equation (NSE) with a saturating nonlinearity.
The saturation function $F(|A|^{2})$ constitutes a "new" type; it
has an unusual form in the sense that in the ultrarelativistic
case $|A|^{2}\gg 1$ it tends to vanish. To the best of our
knowledge NSE with such a "vanishing saturation potential" has not
been derived  and reported so far for any known physical system.

From the refractive index  $\delta n_{nl}=F(I)$ \ ($I=|A|^{2}$ is
the intensity of  the EM field), we may deduce that  the plasma is
self-focusing ($d(\delta n_{nl})/dI>0$) provided $I<1$ while it
becomes defocusing ($d(\delta n_{nl})/dI<0$) for higher
intensities ($I>1$). For a localized intense EM pulse with a peak
intensity $I_{m}>1$, the medium, thus, reacts differently to
different parts of the  EM pulse -- focusing the peak region while
defocusing  the wings.

An immediate consequence of  the  fucusing-defocusing saturating
nonlinearity (originating from the temperature-asymmetry driven
mechanism) can be illustrated by considering a modulation
instability of quasi-monochromatic EM wave. Equation
(\ref{eq:S-6}) is satisfied by the plane wave solution
$A=A_{0}\exp (i\tau F(|A_{0}|^{2}))+c.c.$. The standard stability
analysis then shows that a linear modulation with  frequency
$\Omega $ and  wave number $\mathbf{K}$ obeys the dispersion
relation $\Omega
^{2}=\mathbf{K}^{2}[\mathbf{K}^{2}-2A_{0}^{2}(1-A_{0}^{2})/(1+A_{0}^{2})^{3}]$
which exhibits a purely
growing mode if $A_{0}<1$ and $K<K_{cr}=\sqrt{%
2A_{0}^{2}(1-A_{0}^{2})/(1+A_{0}^{2})^{3}}$, while for the
ultrarelativistic  case ($A_{0}>1$), there is no modulation
instability.  One can expect that the modulation instability of
moderately intense field ($A_{0}<1$)  in the nonlinear stage will
lead to the break up of the field into soliton-like pulses with a
characteristic length corresponding to the optimum scale of
instability ($\sim \sqrt{2}/K_{cr}$).

In what follows we investigate the possibility of finding stable
solitonic solutions of Eq.(\ref{eq:S-6}) under a variety of
conditions. For stationary solitons, we look for solutions  that are
 "spherical" symmetric: $A=A(r)\exp (i\lambda )$ where $\lambda $ is a constant
measuring the nonlinear frequency shift. Notice that the
co-moving coordinate ($\xi $ ) can be treated on an equal footing
with the spatial coordinate ($\mathbf{r}_{\perp }$). In terms of the
radial variable $r=\left( \mathbf{r}_{\perp }^{2}+\xi ^{2}\right)
^{1/2}$, Eq.(\ref{eq:S-6}) reduces to an ordinary differential
equation that  can not be analytically solved. However, to
better understand the results of possible simulations it is helpful  to
rewrite it (after trivial manipulations) as the equation
describing  a "particle" moving with friction in the potential:
\begin{equation}
\frac{d}{dr}\left[ \left( \frac{dA}{dr}\right) ^{2}+V(A)\right]
=-\frac{2(D-1)}{r^{D-1}}\left( \frac{dA}{dr}\right) ^{2}
\label{S-8}
\end{equation}%
where the "effective potential" is $V(A)=-\lambda A^{2}+\ln
(1+A^{2})-A^{2}/(1+A^{2})$. Here $D(=1,2,3)$ dimension of the
problem.

The profile of the potential for different values of the nonlinear
frequency shift $\lambda $ is presented in Fig.1: 1) curve "a" in
this plot corresponds to  $\lambda
>\lambda_{cr}^{(1D)}\simeq 0.2162$, 2) $\lambda
=\lambda _{cr}^{(1D)}$ for the curve "b", and 3) the potential in
the range $0<\lambda <\lambda _{cr}^{(1D)}$ is given in  curve
"c". The solitary solutions correspond  to the effective particle
which at the "moment" $r=0$ rests at a point with coordinate
$A_{m}$. Then it rolls down ($r>0$), dissipates "energy" and
approaches asymptotically ($r\rightarrow \infty $) the potential
maximum at $A=0$. It is obvious that solitary solution can not
exist for $\lambda >\lambda _{cr}^{(1D)}$ \ while for $\lambda
=\lambda _{cr}^{(1D)}$ solitary solution does exist.

However in the range covered in  "c"  ($0<\lambda <\lambda
_{cr}^{(1D)}$) solitary wave solution exists in more than one
dimension; the corresponding $A_{m}$ for a given eigenvalue
$\lambda $  has to  be found numerically. Notice, that in
1-dimension ($1D$), the particle motion becomes conservative
significantly simplifying the problem. By demanding $V(A)=0$ we
find the relationship $\lambda =\ln
(1+A_{m}^{2})/A_{m}^{2}-1/(1+A_{m}^{2}).$ It can be established
analytically that the growing slope of this relation defines the
amplitude of the soliton $A_{m}$. Corresponding nonlinear
dispersion relation $\lambda =\lambda (A_{m})$ is exhibited in
Fig.2 (curve 1D). One can see from these figures that the 1D
solution is restricted from above: \ $A_{m}\leq A_{mcr}\cong
1.4506$ . For small amplitudes  Eq. (\ref{eq:S-6}) reduces to the
standard NSE with a cubic nonlinearity, and 1D soliton solution
can be found analytically (see \cite{bib:MS} for cold plasma case
when $n=n(T)$).

Profiles for large amplitude 1D solitons are exhibited in Fig.3.
One can see, that as $A\to A_{mcr} \ \ (\lambda \to
\lambda_{cr}^{(1D)})$ the profile of the central part of the
soliton flattens and widens at the top. The existence of flat-top
soliton can be explained by the peculiarities of our
focusing-defocusing nonlinearity: the top part of the pulse with
$A>1$ lies in  the defocusing region with a tendency for
diffraction while the wings of the soliton are in the focusing
region preventing the total spread of the pulse. It is interesting
to remark that for $\lambda>\lambda_{cr}^{(1D)}$ the system
supports existence of the dark soliton which is an antisymmetric
function of coordinate with zero intensity at its center. The dark
soliton corresponds to the particle starting at the right maximum
of the curve "c" and going asymptotically towards the left maximum
of the potential. Background intensity of the field $|A_{0}|$ is
bounded from below ($|A_{0}|>A_{mcr}$) but is not restricted from
above. Thus,  a dark soliton with arbitrarily strong background
intensity is possible in 1D. At the critical frequency shift
($\lambda =\lambda _{cr}^{(1D)}$) the dark and bright solitons can
coexist. The coexistence of these solitons is mainly due to the
particular type of nonlinearity encountered in the present model.
Detailed discussion of the properties of dark solitons is beyond
the intended scope of this paper.

\bigskip
In 2D and 3D,  the nonzero "friction force" forces the
corresponding critical values of  $\lambda $  to be less than \
$\lambda _{cr}^{(1D)}$ \ found for the 1D case. In Fig.2, the
relevant curves correspond to numerically obtained dispersion
relations for 2D and 3D, are displayed. Fundamental solitary
solutions  (without zero nodes) are shown in Figures 4 and 5; In
all these examples one fundamental feature of the soliton
persists, namely  that near the critical eigenvalues, the profile
is endowed with  the flat-top shape.

The stability of the obtained solutions can be tested by
applying the Vakhitov \&\ Kolokolov criterion (see \cite{bib:SBM}
and references therein) according to which the solition is stable if \
$\partial N/\partial \lambda >0$ , where \ $N=\int
d\mathbf{r}_{\perp }d\xi \,|A|^{2}$ \ is the soliton energy
("photon number"). We found that in 1D the "photon number"
is always a growing function of $\lambda $ implying that such
solitons are stable against small perturbations.

In Fig.6 we show the dependence of the "photon number" on the
amplitude $A_{m}$ for 2D and 3D solitons. In contrast to the 1D
case, for the soliton to exist in higher dimensions, its energy \
$N$  \ must exceed  a certain critical value \ $N_{cr}$. For a 2D
case, the "photon number"   must exceed the threshold energy \
$N_{cr}=11.6$ for the EM wave to enter the self-guiding solitonic
regime. Since \ $\partial N/\partial A_{m}>0$ , then \ $\partial
N/\partial \lambda >0$ \ because \ $A_{m}$ \ is a growing function
of \ $\lambda $ (see Fig.2) . Thus the 2D soliton with its power
above  the critical power is always stable. In 3D
 $\partial N/\partial A_{m}>0$ provided \ $N>N_{cr}=236.8$ \
and $A_{m}>0.6$.

\bigskip

We have demonstrated that the pair plasmas  with "assymetry" in
initial temperatures  of its constituents can support  stable
large amplitude localized EM wave structures. These structures,
available in arbitrary (1-3)  dimensions, have flat-top shapes for
certain range of parameters. This result is particularly
interesting for laboratory conditions. Since even a "small"
difference (very small!) in temperature may be as effective in
symmetry breaking as the Baryonic correction in early universe,
one could readily engineer the laboratory plasmas to mimic the
cosmic conditions.

According to (\ref{eq:M-40}) the total plasma density in the pulse
localization area increases dramatically for relativistically
intense pulses. Notice, that the general results of the analysis,
as well as Eq. (\ref{eq:S-6}) are valid both for
ultra-relativistic and non-relativistic temperatures, and hence
warrants applications to both astrophysical and laboratory
plasmas.

\section{Mass Asymmetry Between Species}

In this sub-section we explore another obvious source of asymmetry
between species -- a slight \ difference in the masses of positive
and negative-charged particles. This mechanism is different from
the one discussed above; the mass asymmetry is initially given and
is fixed as distinct  from the dynamical asymmetry created due to
temperature differences. Such a plasma can be created by the
injection of appropriate ion-beams into a trap. Electron-hole
plasma in certain semiconductors or e-p collider plasma created by
slightly different Lorentz factor beams\ are also possible
examples of such system. Appropriate conditions for such plasma
production could  readily appear  in Dusty plasmas as well as in
astrophysical Jets, and Pulsar magnetospheres.

\bigskip

Much of the framework for investigating this mechanism has already
been described. For an unmagnetized plasma, one can use the
dimensionless equations (\ref{eq:H-11}) and (\ref{eq:H-12}) in
which  the Lorentz factors of negatively and positively charged
particles are, respectively, changed to $\gamma^{-}=\left[
1+\left( \mathbf{p}^{-}\right) ^{2}\right] ^{1/2}$ and
$\gamma^{+}=\delta^{-1}\left[
\delta^{2}+\left(\mathbf{p}^{+}\right) ^{2}\right] ^{1/2}$, with
$\delta=m^{+}/m^{-}$ being the ratio of positive ($m^{+}$) and
negative ($m^{-}$)\ \ particle masses.

For a circularly polarized EM wave with wave frequency
$\omega_0^{2}=k_0^{2}+(1+\delta^{-1})$ one can readily show that
$\mathbf{P}_{\bot}^{\pm}=\mp\mathbf{A}$ (the equivalent of
(\ref{eq:H-19})). Then the standard treatment  for a transparent
plasma ($\omega_0 \gg (1+\delta^{-1})^{1/2}, \
v_{g}=k_0/\omega_0\simeq 1$) yields the following relations
(equivalent of (\ref{eq:M-2}), (\ref{eq:M-3}) and (\ref{eq:M-5})):
\begin{equation}
\frac{n^{-}}{\gamma^{-}}=\frac{1}{1+\phi} \ , \qquad
\frac{n^{+}}{\gamma^{+}}=\frac{\delta}{\delta -\phi}
\label{eq:b3'}
\end{equation}
and
\begin{equation}
n^{-}=\frac{1}{2}\left[1+\frac{1+|A|^{2}}{(1+\phi)^{2}}\right] , \
n^{+}=\frac{1}{2}\left[
1+\frac{\delta^{2}+|A|^{2}}{(\delta-\phi)^{2}}\right]
\label{eq:b3''}
\end{equation}
leading to
\[
2i\omega\frac{\partial A}{\partial\tau}+\frac{(1+\delta^{-1})}{\omega^{2}}%
\frac{\partial^{2}A}{\partial\xi^{2}}+\nabla_{\perp}^2\,A+
\]
\begin{equation}
+A\phi\left[ \frac{1}{1+\phi}-\frac {1}{\delta\left( \delta-\phi\right) }%
\right] =0 .  \label{b5}
\end{equation}
For small mass differences  $\delta=1+\eta$ (%
$\eta\ll 1$) and  for the characteristic length   $L\gg 1$, the
potential $\phi$ may be calculated to be
\begin{equation}
\phi=\frac{\eta|A|^{2}}{2(1+|A|^{2})}  \label{b7}
\end{equation}
explicitly displaying that $\phi$ is proportional to $\eta$, i.e., $\phi\ll
1 $ for $\eta\ll 1$ (compare with Eqs. (\ref{eq:M-38}) and (\ref{eq:M-43})).
Equations (\ref{b5}) with (\ref{b7}), with appropriate normalization
of the variables and inclusion of transverse field variations, constitute  an
NLSE (\ref{eq:S-6}) with the following saturating nonlinearity
function ($\eta$ will be absorbed in the normalization):
\begin{equation}
F(|A|^{\,2})=1-\frac{1}{(1+|A|^{\,2})^{\,2}} \ .  \label{eq:S-8}
\end{equation}
Notice that this  form of saturation nonlinearity function
coincides with the one obtained in \cite{bib:BM1} for e-p plasma
with a small fraction of heavy ions. The reader may consult Ref.
\cite{bib:BM2}  for detailed analysis of the system; here we will
just summarize the salient features. Equation (\ref{eq:S-6}) (with
(\ref{eq:S-8})) admits a "spherically" symmetric solitary wave
solution. i.e. the "light bullet", a concentration of mass and
energy. And if the "bullet"\ $A_{m}> 0.7$, this "bullet" is
stable. These "light bullets"\ are found to be exceptionally
robust \cite{bib:MBM}: they can emerge from a large variety of
initial field distributions and are remarkably stable. The total
plasma density variation associated with the soliton $\delta n\sim
A^{2}$ is large for $A^{2}\gg1$ ; The solitons with
ultrarelativistic amplitudes create a large concentration of
density.

Thus, the system of pair plasma with slight initial mass asymmetry
between species supports the existence of long lived objects --
light bullets which carry large amounts of mass and energy exactly
the same way as the pair plasmas with small fraction of heavier
ions \cite{bib:BM1,bib:BM2,bib:MS}.

\vspace{0.5cm}

The saturating nonlinearity (\ref{eq:S-7}) caused by an initial
temperature asymmetry seems to be of a new type (vanishing for
intense pulses); it differs quite fundamentally from the one found
in \cite{bib:BM1} (identical to the one originating in a initial
mass asymmetry). One would think that the temperature asymmetry,
manifesting finally as "effective mass" asymmetry, would be
qualitatively similar  to the initial mass asymmetry. But it is
not so. Fortunately we could trace the cause of the difference.

The  first step in the chain does appear to translate the
temperature asymmetry (see (\ref{eq:M-1}))  into a  difference in
the "effective masses" $G^{\pm}m^{\pm}$". But in reality this
"mass asymmetry" is dynamical and dependent on the scalar
potential $\phi$, \ $m_{\mathrm{eff}}^{+}/m_{\mathrm{eff}}^{-}\sim
(1-\epsilon + 0.5\,\nu\,\phi )$ \ while the one originating in an
initial mass asymmetry ($\delta=1+\eta$) is constant in space-time
and never leads to heating/cooling. It is of utmost significance
to realize that the electrostatic potential $\phi$ is important
not only for maintaining the nonlinearity, but also  to create
dynamical temperature asymmetry for hot plasma conditions; it is
the latter consequence that may lead to qualitatively new and
interesting phenomena in such a state of matter.

As a general conclusion we must state that the results of Sec.V
are valid  only for cold plasmas. For initially hot plasmas one
must take into account the temperature inhomogeneity effects
(discussed earlier in the paper) that change the nature of the
nonlinearity and hence the localization characteristics of the EM
waves. However it is worth mentioning, that the main property of
density bunching and energy localization is always there in pair
plasmas with different type initial asymmetries; it is  just the
character of localization that changes with origin of the
asymmetry.

\section{Summary and Conclusions}

\label{sec:conclusions}

The main result of this paper is that a fundamentally new type of
saturating  focusing-defocusing nonlinearity is derived for a
physical system. This composite nonlinearity, originating in a
small temperature asymmetry in the constituent fluids of a pair
plasma, promises the existence of interesting structures that
intense electromagnetic waves can acquire in such plasmas. We have
discussed an illustrative example where different parts of a high
amplitude pulse are effected differently - the  simultaneous
expansion of the peak region and scrunching of the wings  imparts
a flat top shape to the pulse.  Most  consequences of this new
type of nonlinearity are yet to be worked out.

\bigskip

\section*{ACKNOWLEDGEMENTS}

We acknowledge our special debt to the Abdus Salam International Centre for
Theoretical Physics, Trieste, Italy. The work of SMM was supported
by USDOE Contract No.DE-- FG 03-96ER-54366. The work of NLS and VIB was
partially supported by ISTC Project G-1366 and Georgian NSF grant
projects GNSF 69/07 (GNSF/ST06/4-057) and GNSF 195/07
(GNSF/ST07/4-191).


\clearpage

\vspace{2cm}

\begin{figure}
\epsfig{file=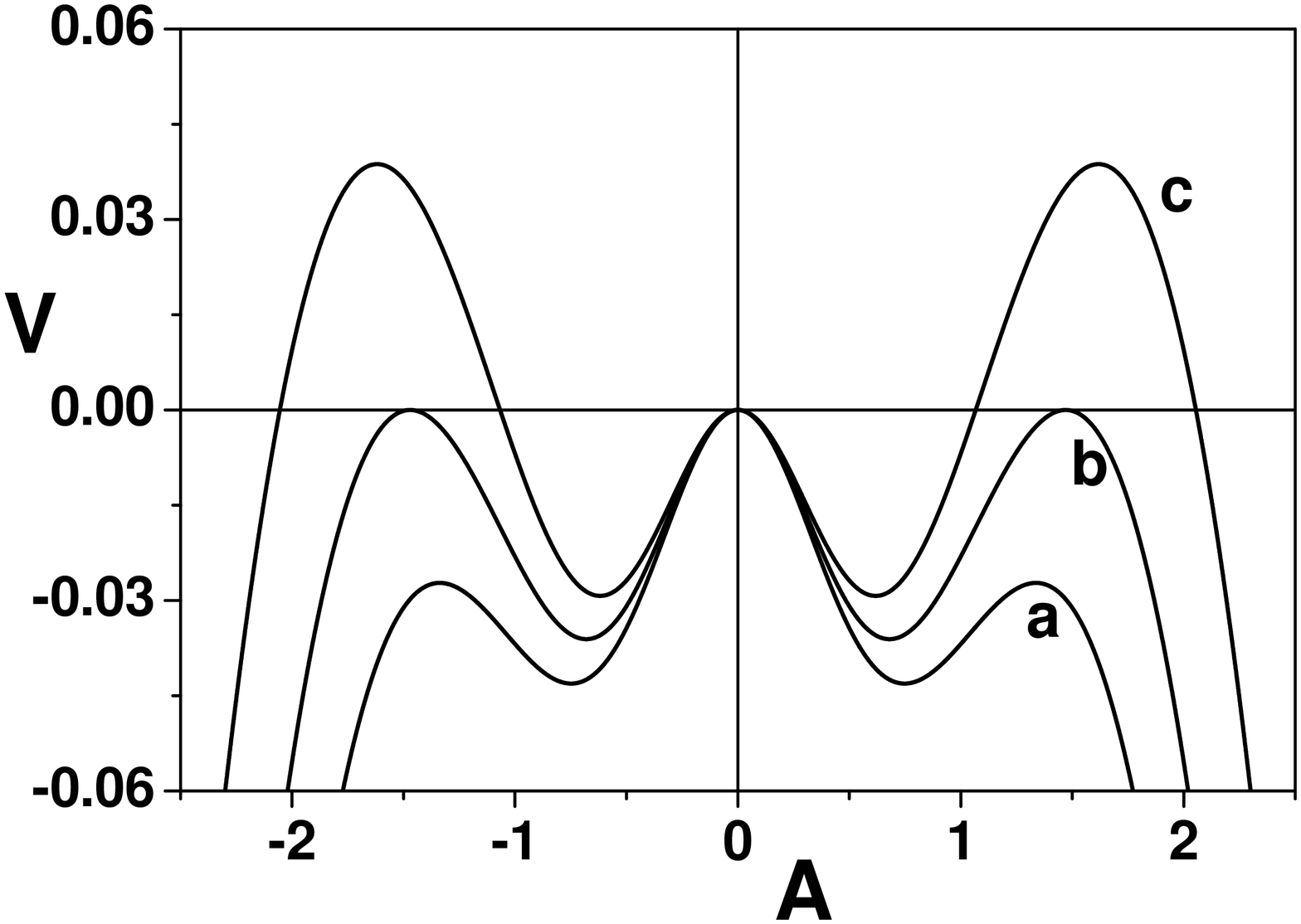,angle=-90,width=4in} \caption{\ \ The
"effective potential" versus the amplitude for different values of
the nonlinear frequency shift $\lambda $. The curve "a"
corresponds to $\lambda >\lambda_{cr}^{(1D)}\simeq 0.2162$, the
curve "b" has $\lambda =\lambda _{cr}^{(1D)}$, and for the curve
"c" $0<\lambda <\lambda _{cr}^{(1D)}$ . }
\end{figure}

\begin{figure}
\epsfig{file=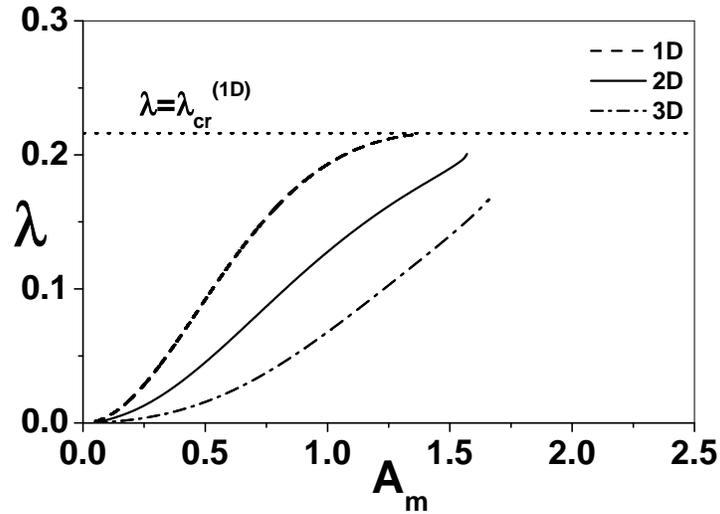,angle=-90,width=4in} \caption{Nonlinear
Dispersion relations: the effective eigenvalue $\lambda $\ as a
function of $A_{m}$. The boundary line (dotted) corresponds to
critical value $\lambda=\lambda_{cr}^{(1D)}$ analytically found
only for 1D. The other three lines represent, respectively the 1D,
2D and 3D dispersion relations.  }
\end{figure}

\begin{figure}
\epsfig{file=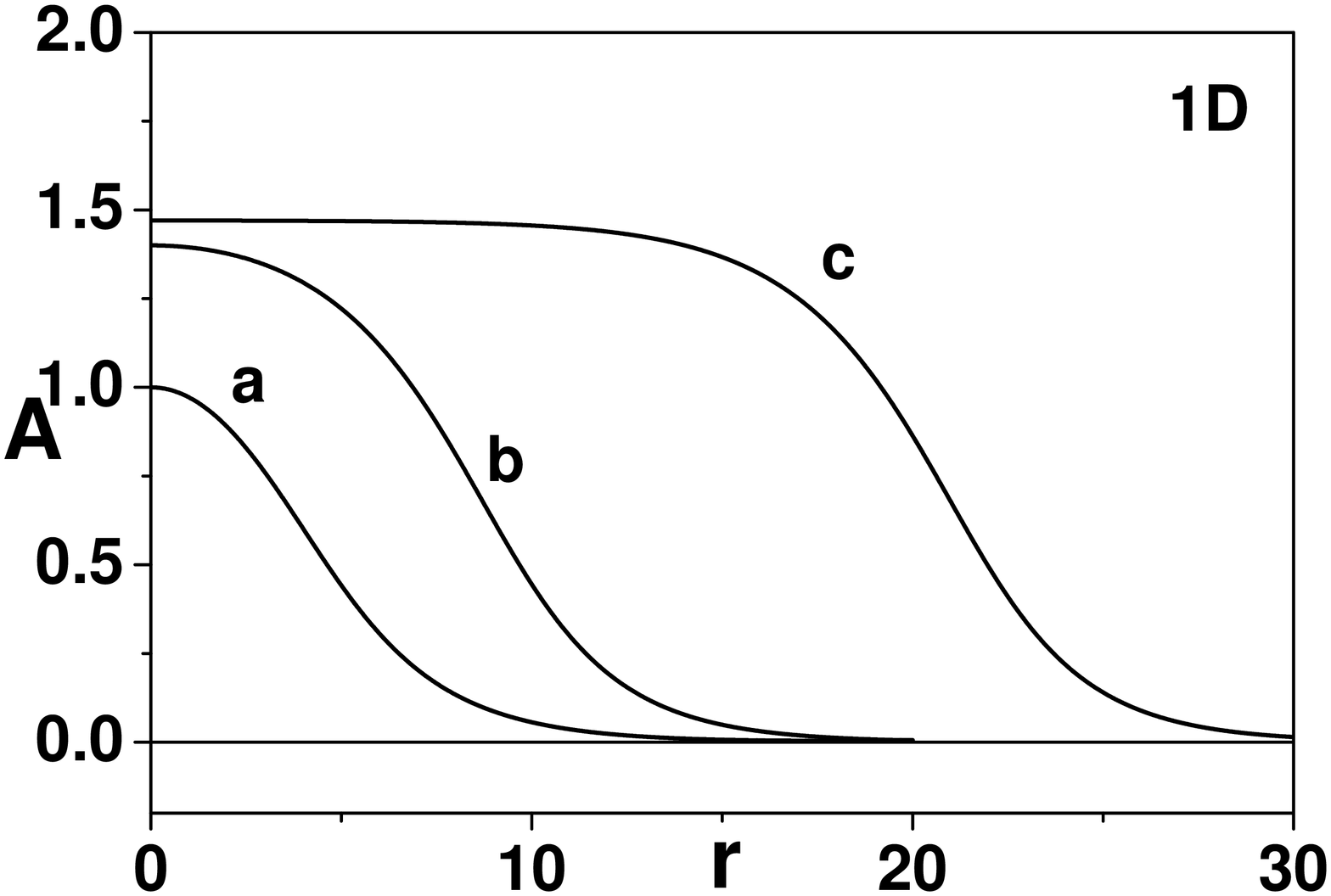,angle=-90,width=4in} \caption{Stationary
soliton solution for 1D for different critical eigenvalues. Plot
"a" corresponds to $\lambda_{cr}=0.19315 \ $ with \ $A_m=1$; plot
"b" corresponds to $\lambda_{cr}=0.21583$ \ with \ $A_m=1.4$ and
plot "c" corresponds to $\lambda_{cr}=0.21622$ \ with \ $A_m=1.47$
respectively. The plot "c" represents the flat-top soliton
solution.}
\end{figure}

\begin{figure}
\epsfig{file=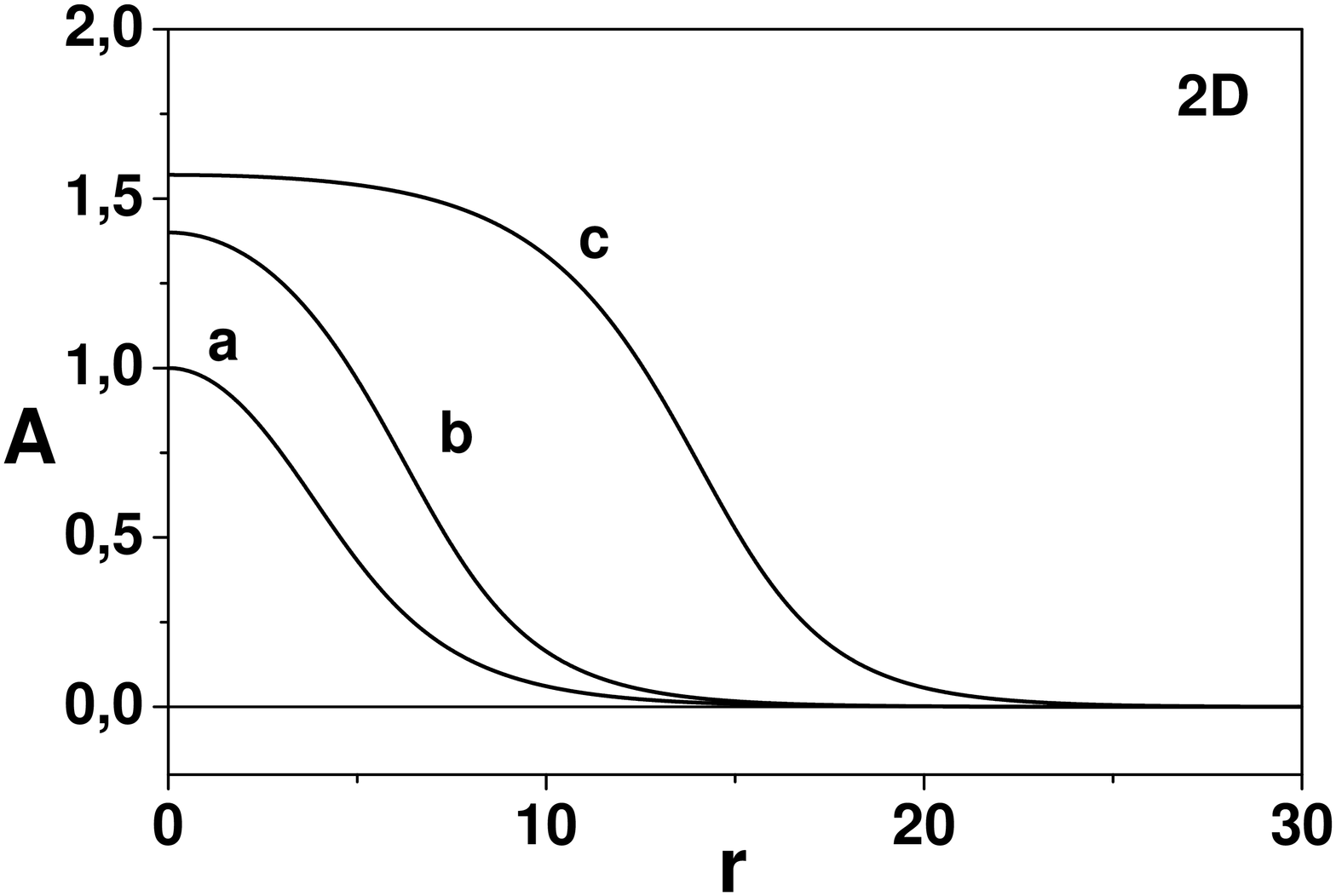,angle=-90,width=4in} \caption{Stationary
soliton solution for 2D for different critical eigenvalues. Plot
"a" corresponds to $\lambda_{cr}=0.12789382$ \ with \ $A_m=1$;
plot "b" corresponds to $\lambda_{cr}=0.17891793$ \ with \
$A_m=1.4$ and plot "c" corresponds to $\lambda_{cr}=0.20299496$ \
with \ $A_m=1.57$ respectively. The plot "c" represents the
flat-top soliton solution.}
\end{figure}

\begin{figure}
\epsfig{file=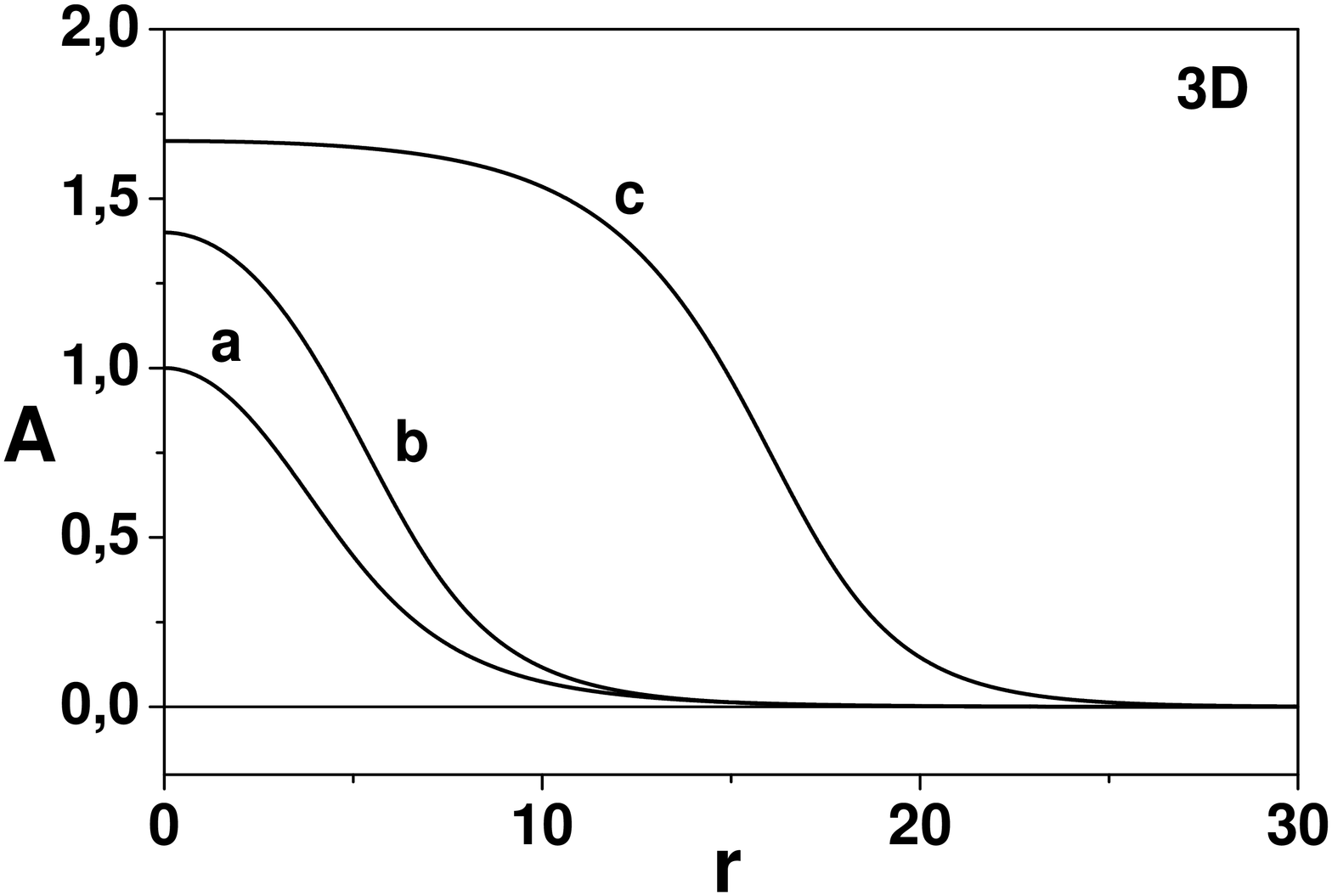,angle=-90,width=4in} \caption{Stationary
soliton solution for 3D for different critical eigenvalues. Plot
"a" corresponds to $\lambda_{cr}=06774722$ \ with \ $A_m=1$; plot
"b" corresponds to $\lambda_{cr}=0.12451945$ \ with \ $A_m=1.4$
and plot "c" corresponds to $\lambda_{cr}=0.19222242$ \ with \
$A_m=1.67$ respectively. The plot "c" represents the flat-top
soliton solution.}
\end{figure}

\begin{figure}
\epsfig{file=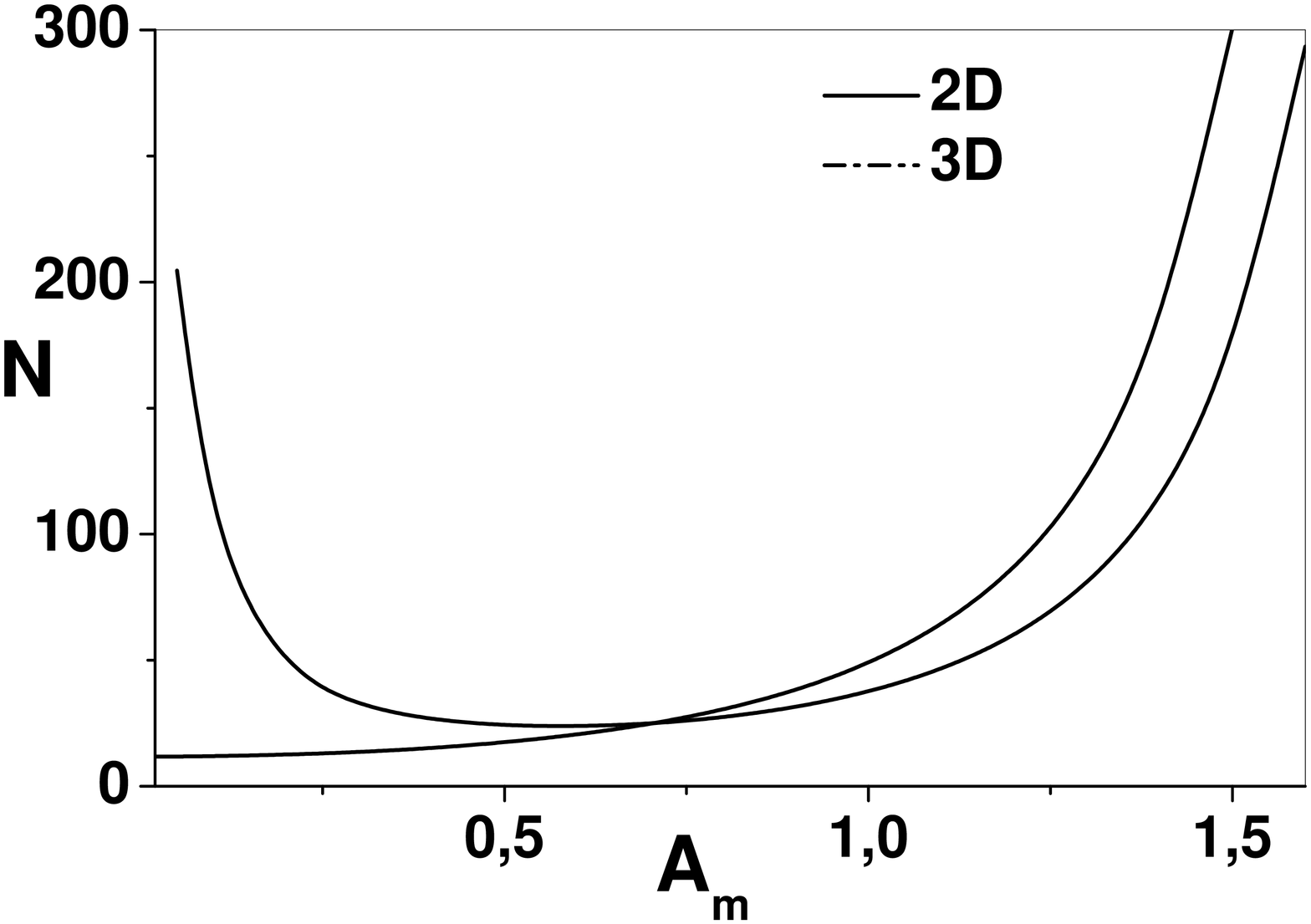,angle=-90,width=4in} \caption{The dependence
of the "photon number" $N$ on the amplitude $A_{m}$ for 2D and 3D.
Normalized "photon numbers" correspond to \ $N^{(2D)}$ (solid
line)\ and \ $10^{-1}N^{(3D)}$ (dashed line). For 2D (3D) the
threshold energy for the existence of soliton is \ $N_{cr}=11.6 \
(236.8) $.}
\end{figure}


\begin{thebibliography}{99}
\bibitem{bib:Wineberg} S. Wineberg. \textit{Gravitation and Cosmology:
Principles and Applications of the General Theory of Relatvity}. John Wiley
and Sons, Inc. 1972.

\bibitem{bib:Rees} M.J. Rees and P. M\'{e}sz\'{a}ros. MNRAS, 258, P41
(1992); P. M\'{e}sz\'{a}ros and M.J. Rees MJ. ApJ 405, 278 (1993); P. M\'{e}%
sz\'{a}ros and M.J. Rees. ApJ. 418, L5 (1993); R. Sari and T. Piran. ApJ
485, 270 (1097)..

\bibitem{bib:blandford} M.C. Begelman, R.D. Blandford, M.J. Rees, Rev. Mod.
Phys. 56 (1984) 255.

\bibitem{bib:sturrock} P.A. Sturrock, Astrophys. J. 164, 529 (1971); M.A.
Ruderman, P.G. Sutherland, Astrophys. J. 196, 51 (1995); F.C. Michel, Theory
of Neutron Star Magnetospheres, University of Chicago Press, Chicago, 1991.

\bibitem{bib:surko} C.M. Surko, M. Leventhal, and A. Passner, Phys. Rev.
Lett. 62, 901 (1989); C. M. Surko and T. J. Murphy, Phys. Fluids B 2, 1372
(1990); R. G. Greaves and C. M. Surko, Phys. Plasmas 4, 1528 (1997).

\bibitem{bib:brams} E.P. Liang, S.C. Wilks, M. Tabak, Phys. Rev. Lett. 81,
4887 (1998); B. Shen, J. Meyer-ter-Vehn, Phys. Rev. E 65, 016405 (2001).

\bibitem{bib:lab-ep} J.W. Shearer et al., Phys. Rev. A 8, 1582 (1973); V.I.
Berezhiani, D.D. Tskhakaya, and P.K. Shukla. Phys. Rev. A, 46, 6608 (1992);
V.I. Berezhiani, D.P. Garuchava and P.K. Shukla. Phys. Lett. A, 360, 624
(2007).

\bibitem{bib:exp-ep} T. E. Cowan et al., Laser Part. Beams 17, 773 (1999);
C. Gahn et al., Appl. Phys. Lett. 77, 2662 (2000); Chen et a., Phys. Rev.
Lett. 102, 105001 (2009).

\bibitem{bib:OH1} W. Oohara and R. Hatakeyama. Phys.Rev.Lett., 91, 205005
(2003); W. Oohara, D. Date, and R. Hatakeyama. Phys. Rev. Lett., 95, 175003
(2005).

\bibitem{bib:OH3} W. Oohara, Y. Kuwabara, and R. Hatakeyama. Phys. Rev. E.,
75, 056403 (2007).

\bibitem{bib:OH4} W. Oohara and R. Hatakeyama. Phys. Plasmas, 14, 055704
(2007).

\bibitem{bib:Ha1} R.Hatakeyama. Private communication (2008).

\bibitem{bib:shukla-ep} M. E. Gedalin, J.G. Lominadze, L. Stenflo and V.N.
Tsitovich, Astrophys. Space Sci. 108, 393 (1985); P.K. Shukla, N.N. Rao,
M.Y. Yu and N.L. Tsintsadze. Phys. Reports. 131, 1 (1986).

\bibitem{bib:theory-ep} T. Cattaert, I. Kourakis and P.K. Shukla. Phys.
Plasmas, 12, 012310 (2005); P. K. Shukla and M. Khan, Phys. Plasmas 2,
014504 (2005); T. Tatsuno, M. Ohhashi V.I. Berezhiani and S.V. Mikeladze.
Phys. Letters, A 363, 225 (2007).

\bibitem{bib:shukla1} B. Eliasson and P. K. Shukla. Phys. Rev. E 71, 046402
(2005).

\bibitem{bib:shukla2} A. Luque, H. Schamel, B. Eliasson and P.K Shukla.
Plasma Phys. Control. Fusion 48, L57 (2006).

\bibitem{bib:TT} T. Tajima and T. Taniuti. Phys. Rev. A, {\bf 42},
3587 (1990); P.K. Shukla, N.L. Tsintsadze and L.N. Tsintsadze.
Phys. Fluids B, {\bf 5}, 233 (1993); D. Farina and S.V. Bulanov.
Phys. Rev E {\bf 64}, 066401 (2001).

\bibitem{bib:hans1} H. Schamel and A. Luque, New J. Phys. 7,69 (2005).

\bibitem{bib:frank} F. Verheest. Phys. Plasmas, 13, 082301 (2006).


\bibitem{bib:H1} H. Saleem. Phys. Plasmas, 13, 044502 (2006);
H. Saleem. Phys. Plasmas, 14, 014505 (2007).

\bibitem{bib:H3} H. Saleem, J. Vranjes, and S. Poedts, Phys. Lett. A A350,
375 (2006).

\bibitem{bib:shukla4} R. Sabry, W.M. Moslem and P.K. Shukla. Phys. Plasmas,
16, 032302 (2009).

\bibitem{bib:H5} H. Saleem, H. and N. Batool. Phys. Plasmas, 16, 022302
(2009).

\bibitem{bib:BM1} V.I. Berezhiani and S.M. Mahajan. Phys. Rev. Lett. 73,
1110 (1994).

\bibitem{bib:BM2} V.I. Berezhiani and S.M. Mahajan. Phys. Rev. E 52, 1968
(1995).

\bibitem{bib:MS} S.M. Mahajan and N.L. Shatashvili. Phys. Plasmas. 15,
100701 (2008).

\bibitem{bib:SJK} N.L. Shatashvili, J.I. Javakhishvili, and H. Kaya.
Astrophys. Space Sci. 250, 109 (1997); N.L. Shatashvili and N. Rao. Phys.
Plasmas, 6, 66 (1999).

\bibitem{bib:shukla3} A. Esfandyari-Kalejahi, I. Kourakis, M. Mehdipoor and
P.K. Shukla. J. Phys. A: Math.Gen., 39, 13817 (2006).

\bibitem{bib:BMYO} V. I. Berezhiani, S. M. Mahajan, Z. Yoshida, and M. Ohhashi.
Phys. Rev. E 65, 047402 (2002).

\bibitem{bib:Chen} X.L. Chen and R.N. Sudan. Phys. Fluids B, {\bf
5},1338 (1993).

\bibitem{bib:SBM} N.G. Vakhitov and A.A. Kolokolov. Izv. Vyssh. Uchebn. Zaved.
Radiofiz. {\bf 16}, 1020 (1973) [Sov. Radiophys. {\bf 9}, 262
(1973)].

\bibitem{bib:MBM} S.M. Mahajan, V.I. Berezhiani and R. Miklaszewki. Phys.
Plasmas. 5, 3264 (1998); K. Hayata and M. Koshiba, J. Appl. Phys. 71, 2526
(1991); N. A. Akhmediev and J. M. Soto-Crespo, Phys. Rev. A 47, 1358 (1993);
V. Skarka, V. I. Berezhiani, and R. Miklaszewski, Phys. Rev. E 56, 1080
(1997).


\end{thebibliography}
\end{document}